\documentclass[amsmath,twocolumn,amssymb,prd,preprintnumbers,showpacs]{revtex4}
\usepackage{amsmath}
\usepackage{amsmath}
\usepackage[cmtip,arrow]{xy}
\usepackage{pb-diagram,pb-xy}
\usepackage{slashed}
\def \be {\begin{equation}}
\def \ee {\end{equation}}
\def \bea {\begin{eqnarray}}
\def \eea {\end{eqnarray}}
\def \sla {\slashed}
\usepackage{graphicx}
\usepackage{amssymb}
\begin{document}
\title{Unitarity in higher-order Lorentz-invariance violating QED}
\author{Carlos M. Reyes}
\email[Electronic mail: ]{creyes@ubiobio.cl}
\affiliation{  Departamento de Ciencias 
B{\'a}sicas, Universidad del B{\'i}o B{\'i}o,\\ Casilla 447, 
Chill\'an, Chile }
%\date{}
\begin{abstract}
The unitarity in Lorentz invariance violating QED consisting of standard fermions and higher-order 
photons of the Myers-Pospelov theory is studied. We find ghost states associated to the 
higher-order character of the theory which could render the $S$-matrix nonunitary.
An explicit calculation to check perturbative unitarity in the process of electron-positron scattering   
is performed and it is found to be possible to preserve unitarity. 
\end{abstract}
%\begin{keyword}
%Effective field theory, Lorentz violation, Higher derivative theory, Causality
%\PACS 11.10.Lm, 11.15.-q, 11.30.Cp 
%\end{keyword}
\maketitle
%-------------------------------------------------------------------------
\section{Introduction}
%-------------------------------------------------------------------------
In recent years, higher-order operators
have become the object of intense study in the search for possible effects of
Lorentz-invariance violation. The consideration of Planck-mass-suppressed higher-order operators 
allows us
to go beyond the limits of renormalizable operators, 
that is, operators with mass dimension of four or less \cite{Colladay-Kostelecky, Data}.

In general, higher-order operators lead to substantial changes of the theory. 
A very special modification occurs when the 
higher-order operators turn into higher-order time derivatives. In this case
the theory involves additional degrees of freedom associated to ultra-high-energy modes. 
Unlike 
what happens with renormalizable operators and their corresponding high-energy modes, the
ultra-high-energy modes are not reduced to the normal ones by setting 
the dimensionless parameters of the effective terms to zero.
These higher-order operators are considered as an effective approach to describe 
an underlying--yet unknown--fundamental theory. It is expected that
the effective approach deviates from the fundamental 
theory at energies of the order of the Planck-mass scale. 
However, far below this scale, the theory might be sensitive to these new effects. 

Many extensions of the standard model have been proposed 
in order to include Lorentz invariance violation using
higher-order operators. For example, they have been proposed for  
 dimension five \cite{M-P,bolokhov} and recently for 
arbitrary dimensions for photons \cite{photons} and fermions \cite{neutrinos}.
They have been studied in loop quantum gravity \cite{loops}, strings \cite{Moeller}
cosmological bounds \cite{grb}, synchrotron radiation \cite{syn}, 
fine-tuning problems \cite{TIMELIKE}, radiative corrections
\cite{rad-corrections},
anisotropies \cite{anisotropies}, causality, and stability \cite{formal}. 
Recently, higher-order operators have received special attention in connection with the 
hierarchy problem in the standard model \cite{BG}. 
Here we are interested in dimension-five operators 
of the photon sector of the Myers and Pospelov theory \cite{M-P}.
In particular, we focus on the study of unitarity in the higher-order QED with standard fermions and
Myers and Pospelov photons.

The organization of the paper is as follows.
In the second section we present the Myers and Pospelov electromagnetic theory and
we study its polarization vectors. We 
obtain the dispersion relations in general backgrounds, giving
 special attention to those cases in which the theory is a 
higher time-derivative theory.  
In the third section we check perturbative unitarity in the 
  electron-positron scattering at tree-level order. Finally, we give 
the conclusions.
%---------------------------------------------------------------------------
\section{The photon Myers and Pospelov model}
%---------------------------------------------------------------------------
The Maxwell-Myers-Pospelov Lagrangian density for photons is given by 
\begin{eqnarray}\label{M-M-P}
\mathcal L_{M.M.P}=-\frac{1}{4}F^{\mu\nu}  F_{\mu\nu}- \frac{\xi}{2M_P} 
n_{\mu}\epsilon^{\mu\nu \lambda \sigma} A_{\nu}(n 
\cdot \partial)^2   F_{\lambda \sigma},
\end{eqnarray}
where $n$ is a four-vector defining a preferred 
reference frame, $M_P$ is the Planck mass, and $\xi$ is a dimensionless parameter. 

The equations of motion derived from the Lagrangian (\ref{M-M-P}) are
\begin{eqnarray}\label{eqmot}
\partial_{\mu} F^{\mu \nu}+g \epsilon ^{\nu \alpha 
\lambda \sigma} n_{\alpha}  (n 
\cdot \partial)^2   F_{\lambda \sigma}=4\pi j^{\nu},
\end{eqnarray}
where we have introduced a source $j^{\nu}$ and defined $g=\xi/M_P$.

In terms of the physical fields
\begin{eqnarray}\label{ELECTRICFIELD}
{\vec E}&=&-\frac{\partial {\vec A}}{\partial t} - \nabla A_0,
\\
{\vec B}&=& \nabla \times {\vec A},
\end{eqnarray} 
we can rewrite Eq.(\ref{eqmot}) as
\begin{eqnarray}\label{ELECTREC1}
&& \nabla \cdot {\vec E}+2g(n\cdot \partial)^2({\vec n}
 \cdot {\vec B})=4\pi \rho, 
\\ &&
-\frac{\partial {\vec E}}{\partial t}+\nabla \times {\vec B}+
2g (n\cdot \partial)^2 (n_0 {\vec B}-({\vec n} 
\times {\vec E}))=4\pi {\vec j}, \label{ELECTREC}\nonumber
\end{eqnarray}
together with the usual identities
\begin{eqnarray}\label{homoge}
  \nabla \cdot {\vec B}&=&0,\nonumber  \\ \nabla 
  \times {\vec E}+ \frac{\partial 
  {\vec B}}{\partial t}&=&0.
\end{eqnarray}
It can be shown that the conserved  
energy-momentum tensor is given by
\begin{eqnarray}
T^{\alpha}_{\;\;\beta}=-G^{\alpha \gamma}
F_{\beta \gamma} -
\delta^{\alpha}_{\beta} \mathcal L+\frac{1}{2}
(\partial_{\beta}A_{\gamma} 
-A_{\gamma} \partial_{\beta}) G^{\alpha \gamma},
\end{eqnarray}
where 
\begin{eqnarray}
G^{\mu\nu }=F^{\mu\nu }+2g \epsilon ^{\mu \nu \alpha 
\beta} n_{\alpha}  (n 
\cdot \partial)^2   A_{\beta},
\end{eqnarray} 
Now, making an analogy with electrodynamics in macroscopic media, 
we can define an effective vector displacement field 
${ \vec D}$  and an effective pseudovector magnetic field
${\vec  H}$ \cite{photons},
\begin{eqnarray}\label{macroscopic}
 \vec D  &\equiv &    \vec E-2g (n\cdot \partial)^2
 (\vec n \times \vec A),\nonumber 
\\
 \vec  H  &\equiv&   
  \vec B + 2g 
(n\cdot \partial)^2 ( n_0\vec A- \vec n A_0),
\end{eqnarray} 
such that $ D^i=G_{0i}$ and $-\epsilon^{ijk} H^k=G_{ij}$.
In terms of these fields, the energy and momentum density are
\begin{eqnarray}
\mathcal H&=&\frac{1}{2}(D\cdot E+B\cdot H)+
\frac{1}{2}\left( \frac{\partial \vec A}
{\partial t} \cdot \vec D  -\vec A \cdot \frac
{\partial \vec D}{\partial t}\right),
\nonumber \\
\vec {\mathcal S} &=&\vec D\times \vec B+\frac
{1}{2}\left(\vec A\cdot \nabla\vec 
D-\nabla (\vec A\cdot \vec D)  \right),
\end{eqnarray}
where $\mathcal H=T^{00}$ and ${\mathcal S}^i =T^{0i}$.
%---------------------------------------------------------------------------
\subsection{Polarization vectors and dispersion relations}
%---------------------------------------------------------------------------
In this subsection, we cast the Myers-Pospelov model in terms of 
a basis of
four-vectors analogous to the left- and right-handed polarizations of 
usual electrodynamics. This will allow 
us to find the dispersion relation in an easier way
and to simplify the analysis of unitarity in the next section. 

Our first task is to take advantage of the
similar Lagrangian structures of
the Myers-Pospelov and Chern-Simons theories, recalling that they only 
differ by the inclusion of the operator $(n\cdot \partial)^2$.
Hence, let us start with the tensor 
\begin{eqnarray}
e^{\mu\nu}=\eta^{\mu \nu}-\frac{(n\cdot k)}{D}
(n^{\mu}k^{\nu}+n^{\mu}k^{\nu})+\frac{k^2}{D} n^{\mu}n^{\nu}
+\frac{n^2}{D} k^{\mu}k^{\nu},\nonumber \\
\end{eqnarray}
 and the pseudotensor
\begin{eqnarray}
\epsilon^{\mu \nu}=D^{-1/2}\epsilon^{\mu \alpha \rho \nu}n_{\alpha}k_{\rho},
\end{eqnarray}
where $D(k,n)=(n\cdot k)^2-n^2k^2$, see Refs. 
\cite{cher-s,Andrianov}.

Both quantities $e^{\mu\nu}$ and $\epsilon^{\mu \nu}$ can be considered  
projectors onto the 
two-dimensional hyperplane orthogonal to $k^{\mu}$ and $n^{\mu}$.
Indeed, it can be verified that when the preferred 
four-vector is purely timelike,
the tensor $e^{\mu\nu}$
reduces to the transverse delta $\delta^T_{ij}=\delta_{ij}-
\frac{k_i k_j}{|\vec k|^2}$. Also, 
it is straightforward to check that both tensors satisfy
the transverse relations $e^{\mu\nu}n_{\nu}=e^{\mu\nu}
k_{\nu}=\epsilon^{\mu \nu}n_{\nu}=\epsilon^{\mu \nu}k_{\nu}=0$. 

Now, choosing a frame on the two-dimensional 
hyperplane, we can always select a real basis
of four-vectors $e_{\mu}^{(a)}$ to be orthonormal 
\begin{eqnarray}
 \eta^{\mu \nu} e_{\mu}^{(a)}e_{\nu}^{(b)}=-\delta^{ab},
\end{eqnarray}
and to have the properties
\begin{eqnarray}\label{polsum}
 e_{\mu\nu}&=&-\sum_{a=1,2} e_{\mu}^{(a)}e_{\nu}^{(a)},
\\
\epsilon^{\mu \nu}&=&e^{(1)\mu} e^{(2)\nu}-e^{(2)\mu}e^{(1)\nu}.
\end{eqnarray}
We can switch to a basis of complex polarization four-vectors defining
\begin{eqnarray}\label{pol-vect}
\varepsilon^{\lambda}_{\mu}
 =\frac{1}{\sqrt{2}}(e_{\mu}^{(1)}+i\lambda \,e_{\mu}^{(2)}),
\end{eqnarray}
where $\lambda =\pm$.
It can be checked that any four-vector $J_{\mu}$ 
can be decomposed in this basis as
\begin{eqnarray}
J^{+}_{\mu}&=&P^{+}_{\mu \nu} J^{\nu},
\end{eqnarray}
and
\begin{eqnarray}
J^{-}_{\mu}&=&P^{-}_{\mu \nu} J^{\nu},
\end{eqnarray}
where the orthogonal projectors $P^{\lambda}_{\mu \nu}$ 
are defined by
\begin{eqnarray}\label{projector}
P^{\lambda}_{\mu \nu}= \frac{1}{2} (e_{\mu \nu}+i\lambda 
 \epsilon_{\mu \nu}).
\end{eqnarray}
Some useful properties are
\begin{eqnarray}
\epsilon^{\mu \nu} e^{(1)}_{\nu}= e^{(2)\mu}, \qquad  
\epsilon^{\mu \nu} e^{(2)}_{\nu}=- e^{(1)\mu},
\end{eqnarray}
\begin{eqnarray}
\epsilon^{\mu \alpha} \epsilon^{ \nu}_{\;\;\alpha}=e^{\mu \nu},
\qquad
\epsilon^{\nu \mu }=  e^{ \nu \alpha} \epsilon ^{\mu }_{\;\;\alpha},
\end{eqnarray}
\begin{eqnarray}\label{imp-rel}
 P^{\lambda}_{\mu \nu}=- \varepsilon^{\lambda}_{\mu}
\varepsilon^{*\lambda}_{\nu}.
\end{eqnarray}
Now, consider the gauge field expanded in terms of the new basis as
\begin{eqnarray}
A_{\mu}(x)=\sum_{\lambda}\int d^3 k \left(\widetilde A^{\lambda}(k) 
\varepsilon_{\mu}^{\lambda}(k)  e^{-ik\cdot x} 
+\widetilde A^{*\lambda}
(k) \varepsilon_{\mu}^{*\lambda}(k)  e^{ik\cdot x}\right).\nonumber \\
\end{eqnarray}
Replacing this in the equation of motion (\ref{eqmot}), we arrive at
\begin{eqnarray}\label{propagator}
 \left( \begin{array}{cc}
(G^{+})^{-1} & 0 \\ 0 & (G^{-})^{-1}  \end{array} \right)
 \left( \begin{array}{c} 
\widetilde A^{+} \\ \widetilde A^{-} \end{array} \right) 
=4\pi \left( \begin{array}{c} 
 j^{+} \\  j^{-} \end{array} \right),
\end{eqnarray}
where $(G^{\lambda})^{-1}  = (k^2+2g\lambda
 (n\cdot k)^2{\sqrt{D}})$.
Solving the $2\times 2$ determinant, the dispersion relation reads
\begin{eqnarray}\label{DR}
G= (k^2)^2-4g^2(n \cdot k)^4 \left((n\cdot k)^2-n^2k^2\right)=0,
\end{eqnarray}
in agreement with the work in \cite{formal}.
%----------------------------------------------------------------------------------------
\subsection{Minimal extensions}
%-----------------------------------------------------------------------------------------
The Myers-Pospelov theory can be defined in certain 
backgrounds, where the modifications 
are perturbative connected to the usual theory.
The new physics includes
 birefringence, anisotropies, and modified dispersion relations, which is proper 
in the Lorentz symmetry breakdown \cite{sme1,sme2,sme3}.
However, its degrees of freedom are not increased 
compared to the standard field theory. Moreover, we can always 
reobtain the usual theory by 
taking the appropriate low energy-limit.
As we will see later, in more general backgrounds the theory 
allows us to produce additional 
degrees of freedom associated to ghost states whose frequency 
solutions diverge when taking the limit $g\to 0$.

There are two possible ways to define the theory minimally.
The first one is
to choose a purely timelike background $n=(1,0,0,0)$ for which 
the positive solutions are found to be
\begin{eqnarray}
\omega^{\lambda}_{T}=\frac{ | \vec k |}{\sqrt{1+2g\lambda{ | \vec k |}}},
\end{eqnarray}
where $\lambda$ labels the circular polarization vectors 
introduced earlier. 
It is clear that the solution $\omega_T^{-}$ remains 
real only in the region defined by $| {\vec k}|<1/(2g)$.
For higher momenta the negative mode becomes complex, 
introducing instabilities in the theory.
If one restricts to real solutions then the corresponding Feynman diagrams 
will depend on a natural cutoff having the possibility
to introduce fine-tuning effects \cite{f-t} and unitarity 
violation \cite{Lee-Wick}. Some methods have been 
investigated in order to avoid the fine-tuning problem \cite{TIMELIKE}.

The second possibility is to consider a purely spacetime background.  
In this case the dispersion relations reads
\begin{eqnarray}
\omega^{\lambda}_{S}&=&\left(k^2+2g^2(n\cdot k)^4  +\nonumber \right. \\&& \left.
\lambda\left( n\cdot k  \right)^3 (1+g^2\vec n^4(\vec n\cdot \vec k)^2
 )^{1/2}\right)^{1/2}.
\end{eqnarray}
By simple inspection one can see 
that the solutions are always real.
The spacelike case has been discussed in 
relation to anisotropies \cite{anisotropies} and microcausality \cite{formal}. 
%---------------------------------------------------------------------
\subsection{Higher-order sector}
%---------------------------------------------------------------------
The variational formalism for higher-order time derivative theories
was developed some time ago by Ostrogradski \cite{ostro}.
 Since then, these theories 
 have been studied in 
different contexts.
In quantum field theory, higher-order time derivatives are attractive 
since they can improve the ultraviolet properties of the theory
\cite{Pod,ult-div}. Unfortunately, they also 
introduce negative norm states or ghosts which may destroy the probabilistic 
interpretation \cite{p-u}. 

Lee and Wick studied an equivalent description 
to higher-order theories based on
indefinite metrics in Hilbert space \cite{Lee-Wick}.
In many cases 
one can find explicitly the equivalence by performing a transformation on the basic variables
of the higher-order theory. 
The transformation takes
 the higher-order Lagrangian
into a sum of two normal-order Lagrangians but with one having  
a minus sign in front.
It was shown that, in effect, the ghosts that appear can lead to the loss of unitarity.
However, by demanding ghost particles
to be unstable and thus not be asymptotic states,
they were able to show 
that it was possible to define a unitary $S$-matrix. Also, Cutkosky
used a generalized cutting rule framework to prove that 
unitarity can be preserved in a general class of diagrams \cite{CUTW}.
Both prescriptions, however, were shown to introduce noncausal effects.

To see how the ghosts appear in our model, 
let us write the propagator as
\begin{eqnarray}
D_{\mu \nu}=\frac{d_{\mu \nu}}{(k^2)^2-4g^2(n \cdot k)^4 
\left((n\cdot k)^2-n^2k^2\right)},
\end{eqnarray}
in accordance with the pole structure previously found. Here  
 $d_{\mu \nu}$ is some tensor, which for the moment we can ignore.
Let us focus on the denominator, 
\begin{eqnarray}
\frac{1}{G}=\frac{1}{(k^2)^2-4g^2(n \cdot k)^4 \left((n\cdot k)^2-n^2k^2\right)},
\end{eqnarray}
which can be rewritten in Euclidean space with $x_0\to ix_{0E}$, as 
\begin{eqnarray}\label{poles}
\frac{1}{G_E}=\frac{1}{k^2_E}-\frac{4g^2   (n^2_E)^3\cos^4 
\theta \sin^2 \theta }{1+4g^2 k_E^2 (n^2_E)^3\cos^4 \theta \sin^2 \theta},
\end{eqnarray}
where $\theta$ is the angle between $k_E$ and $n_E$ and the notation 
is $x_E=(x_{0E},\vec x)$. 

Now the following is clear:

 (i) There is an additional pole in the second term of the 
right-hand side of Eq. (\ref{poles}) given by
the solution of $1+4g^2 k_E^2 (n^2_E)^3\cos^4 \theta \sin^2 
\theta=0$, besides the usual one in the first term
$k_E^2=0$. 

(ii) This extra pole produces a negative residue contribution,
which is interpreted as a negative norm state particle or ghost state \cite{bender}.

Having identified the ghost contribution, let us analyze with more detail 
the higher-order 
sector. We focus on the lightlike case $n^2=0$, 
where the dispersion relation is simplified. Other cases imply solving
sixth-order algebraic equations that may be tedious and do not 
contribute decisively to the discussion.

The dispersion 
relation in this case is
\begin{eqnarray}\label{part}
(G^{\lambda})^{-1}=\omega^2- {\vec k}^2+2g\lambda(n_0 
\omega  - \vec n\cdot \vec k )^3=0.
\end{eqnarray}
Note that given the form of the above equation
we do not have positive and negative energy 
solutions as in the pure time and space cases. 
Instead from (\ref{part}), we have the relation 
\begin{eqnarray}\label{trans}
-\omega_{[i]}^{\lambda}(-\vec k)=\omega_{[i]}^{-\lambda}(\vec k),
\end{eqnarray}
where the index runs over $i=0,1,2$ and $[i]$ 
denotes any of the three solutions for each $\lambda$.
Without loss of generality we can consider $n=(1,0,0,1)$ in which case 
the exact solutions are
\vspace{-20pt}
\begin{widetext}
\begin{eqnarray}\label{higher-order-freq}
\omega^{\lambda}_0&=&-\frac{1-6g\lambda k_z}{6g\lambda}-
\frac{-1+12g\lambda k_z}{3 \times2^{2/3}g\lambda \Delta^{\lambda\,}}
+\frac{\Delta^{\lambda \,}}{6 \times2^{1/3}g\lambda},
\\
\omega^{\lambda}_1&=&- \frac{1-6g\lambda k_z}{6g\lambda}+\frac{(1+i\sqrt{3})
(-1+12g\lambda k_z)}{6 \times2^{2/3}g\lambda
 \Delta^{\lambda\,}}-\frac{(1-i\sqrt{3}) 
\Delta^{\lambda \,}}{12 \times2^{1/3}g\lambda},\nonumber 
\\
\omega^{\lambda}_2&=& - \frac{1-6g\lambda k_z}{6g\lambda}+
\frac{(1-i\sqrt{3})(-1+12g\lambda 
k_z)}{6 \times2^{2/3}g \lambda \Delta^{\lambda\,}}-
\frac{(1+i\sqrt{3}) \Delta^{\lambda\,}}{12 
\times2^{1/3}g\lambda},\nonumber 
\end{eqnarray}
where
\begin{eqnarray}
\Delta^{\lambda\,}=\left( -2+108g^2 \vec k^2
+36g\lambda k_z-108g^2k_z^2   
+\sqrt{ (-2+108 g^2 \vec k^2+36g\lambda k_z-108g^2k_z^2)^2 
 +4(-1+12g\lambda k_z)^3 } \right)^{1/3}. 
\end{eqnarray}
\end{widetext}
Under the transformation (\ref{trans}) we note that 
\begin{eqnarray}
\omega^{+}_0 \to \omega^{-}_0,   \qquad 
\omega^{+}_1\to \omega^{-}_2, \qquad  
\omega^{+}_2\to \omega^{-}_1.
\end{eqnarray}
The approximations for small $g$ are
\begin{eqnarray}\label{approximated}
\omega^{\lambda}_0&\approx&-\frac{1}{2g\lambda }
+3 k_z+2g\lambda(\vec k^2+3 k_z^2),
\\
\omega^{\lambda}_1&\approx& | \vec k|
-\frac{g\lambda(|\vec k|- k_z)^3}{|\vec k|},
\\
\omega^{\lambda}_2&\approx& - |\vec k|
-\frac{g\lambda(|\vec k|+ k_z)^3}{|\vec k|}.
\end{eqnarray}
 We see that the first solution or the ghost mode
goes to infinity in the limit $g\to 0$, while 
the other two behave as perturbative corrections in the same limit.
%--------------------------------------------------------------------------
\section{Perturbative unitarity}
%---------------------------------------------------------------------------
In this section we study the unitarity in the QED theory 
consisting of higher-order photons of 
Myers-Pospelov minimally coupled to standard fermions.
We verify perturbative unitarity checking the optical theorem 
in the process of electron-positron scattering at tree-level order. 
%--------------------------------------------------------------------------
\subsection{The optical theorem}
%---------------------------------------------------------------------------
The mathematical statement of conservation of the total probability,
 for an arbitrary final state to arise from some initial state in a scattering process, gives 
the unitarity property of the $S$ matrix
\begin{eqnarray}\label{uni}
S^{\dag}S=1.
\end{eqnarray}
The optical theorem relates 
the imaginary part of the forward scattering amplitude 
to the total cross section
and follows from this conservation of probability.
To see how the optical theorem appears, let us consider 
the $S$-matrix in the form,
\begin{eqnarray}
S=1+iT.
\end{eqnarray}
Substitution in Eq. (\ref{uni}) implies the equation 
\begin{eqnarray}
-i(T-T^\dag )= T^{\dag}T.
\end{eqnarray}
Taking the matrix elements 
between initial $ \left| i \right\rangle$ and final $ \left\langle
 f\right|$ states, we have
\begin{eqnarray}
\left\langle f\right| T\left| i \right\rangle -\left\langle f\right|
 T^\dag \left| i \right\rangle =i \left\langle f\right|
 T^{\dag}T  \left| i \right\rangle.
\end{eqnarray}
Now, inserting a complete set of intermediate states 
$\left\langle m\right|$, we rewrite the above equation as
\begin{eqnarray}
\left\langle f\right| T \left| i \right\rangle -
\left\langle f\right|
 T^\dag \left| i \right\rangle =i \sum_{m} \int 
d\Pi_m \left\langle f\right| T^{\dag}\left| m \right\rangle 
\left\langle m\right|  T  \left| i \right\rangle.
\end{eqnarray}
By defining 
\begin{eqnarray}
 \left\langle  f \right|  T  \left|i \right\rangle=\mathcal
 M_{fi} (2\pi)^4 \delta^ {4}(P_f-P_i ),
\\
 \left\langle  f \right|  T^{\dag}  \left|i \right\rangle=
\mathcal M^*_{if}  (2\pi)^4\delta^ {4}(P_f-P_i ),
\end{eqnarray}
we write
\begin{eqnarray}
  \mathcal M_{fi}-\mathcal M^*_{if}= i\sum_{m} \int d\Pi_m 
 \mathcal M_{fm} \mathcal M_{im}^{*},
\end{eqnarray}
and in the special 
case of forward scattering $f=i$, we arrive at the unitarity condition
\begin{eqnarray}\label{unitarity}
2 \,{\rm {Im}} \mathcal M_{ii}= \sum_{m} \int d\Pi_m
\left| \mathcal M_{im} \right|^2,
\end{eqnarray}
where the sum runs over all intermediate states
 that are allowed by the conservation of
total energy and momentum.
Any violation of unitarity due to Lee-Wick ghost fields is expected to  
show up as a contradiction of this unitarity condition constraint
 equation (\ref{unitarity}).
The generalization of the optical theorem 
for Feynman diagrams has been proven by Cutkosky 
using a set of cutting rules \cite{cutkosky2}. 
%--------------------------------------------------------------------------
\subsection{The modified QED}
%---------------------------------------------------------------------------
%%%%%%%%%%%%%%%%%%%%%%%%%%%%%%%%%%%%%%%%%%%%%%%%%%%%%%%%%%%%%%%
\begin{figure*}  [t] 
\includegraphics[
width=0.7\textwidth]{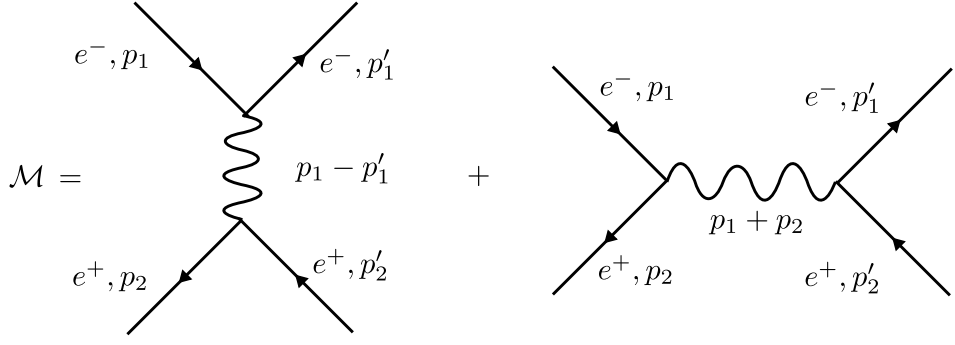}
\caption{\label{Fig3} Direct and exchange graphs contributing 
to the scattering amplitude $\mathcal M(e^+e^-\to e^+e^-)$.  }
\end{figure*}
%%%%%%%%%%%%%%%%%%%%%%%%%%%%%%%%%%%%%%%%%%%%%%%%%%%%%%%%%%%%%%%%
Let us consider the QED Lagrangian
\begin{eqnarray}
 \mathcal L =\bar \psi(i\sla{\partial}-m)\psi-\frac{1}{4}F^{\mu\nu}
  F_{\mu\nu}\nonumber \\-\frac{g}{2} 
n_{\mu}\epsilon^{\mu\nu \lambda \sigma} A_{\nu}(n 
\cdot \partial)^2   F_{\lambda \sigma}+\mathcal L_{int}, 
\end{eqnarray}
where the interaction term $\mathcal L_{int}$ in 
principle, can receive contributions from the dimension-five operators.
These additional terms proportional to 
$(n \cdot A)^2$ coming from the minimal substitution can 
introduce additional vertices that may have to be included in the analysis.
In the transverse gauge, however, we have simply
\begin{eqnarray}
 \mathcal L_{int}=-e 
\bar \psi \gamma^{\mu} A_{\mu}\psi.  
\end{eqnarray}
We will consider the tree-order amplitude of the Bhabba scattering
$e^+e^-\to e^+e^-$ shown in Fig. 1. 
Let us start with the left-hand side of the unitarity condition (\ref{unitarity}). 
The amplitudes that contribute to the $S$ matrix are the direct graph
\begin{eqnarray}\label{ampdir}
\mathcal M^{{\rm dir }}=  (-ie)^2 \int d^4 k\, \delta^{4} 
(p_1-p_1^{\prime}-k) \widehat U^{\mu} U^{\nu} G_{ \mu \nu }(k),
\end{eqnarray}
and the exchange graph
\begin{eqnarray}\label{amplitud2}
\mathcal M^{{\rm ex }}=  (-ie)^2 \int d^4 k\, \delta^{4}
 (p_1+p_2-k) \widehat V^{ \mu} V^{\nu} G_{ \mu \nu }(k),
\end{eqnarray}
where $\widehat U^{ \mu}=  N_{p_2} N_{p_2^{\prime}} \bar v(p_2) 
\gamma^{\mu}v(p_2^{\prime})$, $U^{\nu}= N_{p_1^{\prime}} N_{p_1} 
\bar u(p_1^{\prime}) \gamma^{\nu}u(p_1)$ and 
$\widehat V^{ \mu}= N_{p_1^{\prime}} N_{p_2^{\prime}} \bar
 u(p_1^{\prime}) \gamma^{\mu}v(p^{\prime}_2)$, $V^{\nu}=
 N_{p_2} N_{p_1}
\bar v(p_2) \gamma^{\nu}u(p_1)$, and where $N_{p}=
\sqrt{ \frac{m}{E_{p}}}$ are the usual fermionic normalization constants.

It is not difficult to see that the photon propagator 
is $ G_{ \mu \nu }(k)=-\sum_{\lambda}
 P^{\lambda}_{\mu \nu}G^{\lambda}$,
where the projector is given in (\ref{projector}). 
To simplify we will consider 
the lightlike case where we have a ghost state with frequencies 
$\omega_0^{\pm}$ and two photons with frequencies $\omega_{1,2}^{\lambda}$ 
given in (\ref{higher-order-freq}). 
The propagator in the lightlike case is 
\begin{eqnarray}\label{PROPAGATOR}
G_{ \mu \nu }(k)&=&-\sum_{\lambda} 
\frac{P^{\lambda}_{\mu \nu}(k)}{ k^2+2g\lambda(n\cdot k)^3+
i\epsilon  },
\end{eqnarray}
where and we have included the $i\epsilon$ prescription.

We are interested in the imaginary part of the 
forward-scattering amplitude; therefore, let us set
$p_1^{\prime}\to p_1$ and $p_2^{\prime}\to p_2$. 
Moreover, we can see that the direct process does not 
contribute since the virtual photon can never be on shell
for nonzero external momenta, hence ${\rm Im}[\mathcal M^{{\rm dir }}]=0$.
Let us find the contribution of the exchange process 
and substitute the propagator (\ref{PROPAGATOR}) in (\ref{amplitud2}),
\begin{eqnarray}
\mathcal M^{{\rm ex }}&=&  e^2  \int \frac{d k^0}{(2\pi)}
\int \frac{d^3 \vec k}{(2\pi)^3}\delta^4(p_1+p_2-k)  
V^{ \mu} V^{* \nu}\nonumber \\
&&\times
\sum_{\lambda} \frac{P^{\lambda}_{\mu \nu}(k)}{ 
 k^2+2g\lambda(n\cdot k)^3+i\epsilon  }     .
\end{eqnarray}
 Since only the poles can contribute to the
 imaginary part, it is convenient to rewrite the propagator as
\begin{eqnarray}\label{delta}
&&\frac{1}{  k^2+2g\lambda(n\cdot k)^3+i\epsilon  }  \\
&& 
= \frac{ k^2+2g\lambda(n\cdot k)^3 }{(k^2+2g
\lambda(n\cdot k)^3)^2+\epsilon^2}
-\frac{i\pi \delta (k_0-\omega_1^{\lambda})}
{ 2g\lambda (k_0-\omega_0^{\lambda}) 
(k_0-\omega_2^{\lambda})},\nonumber 
\end{eqnarray}
where we have used the identity $\pi\delta(x)=
\frac{\epsilon}{x^2+\epsilon^2}$, $\epsilon\to 0^+$.

Because of energy conservation encoded 
in $\delta^4(p_1+p_2-k)$, we have that only the positive poles of the virtual 
photon have a chance to contribute. 
We can discard the ghost contribution since its
 energy $| \omega_0 ^{\lambda}| \sim 1/2g$ lies 
beyond the region of validity of the effective theory. 
That is, the external fermions
will always fulfill the condition $p_{01}+p_{02}<|\omega_0^{\lambda}|$. 

Considering (\ref{delta}), we have
\begin{eqnarray}
2{\rm Im}[\mathcal M^{{\rm ex }}]&=&- e^2  \int d k^0
\int \frac{d^3 \vec k}{(2\pi)^3}\delta^4(p_1+p_2-k)        
 V^{\mu} V^{* \nu}\nonumber \\
&&\times  \sum_{\lambda}\frac{  
 P^{\lambda}_{\mu \nu}  \delta (k_0-\omega_1^{\lambda})}
{2g\lambda (k_0-\omega_0^{\lambda}) (k_0-
\omega_2^{\lambda}) }, \nonumber
\\ 
&=& -e^2
\int \frac{d^3 k}{(2\pi)^3}\delta^4(p_1+p_2-k)  
    V^{\mu}  V^{* \nu} \nonumber \\
&&\times \sum_{\lambda}
\frac{P^{\lambda}_{\mu \nu}}{2g\lambda 
(\omega_1^{\lambda}-\omega_0^{\lambda}) 
(\omega_1^{\lambda}-\omega_2^{\lambda}) },   \nonumber \\
&=& e^2\int \frac{d^3 k}{(2\pi)^3   }\delta^4(p_1+p_2-k)  
   V^{\mu}   V^{* \nu}\nonumber \\
&&\times \sum_{\lambda} \frac{
 \varepsilon_{\mu}^{\lambda}  \varepsilon_{\nu}^{*\lambda} }{2g\lambda 
 (\omega_1^{\lambda}-\omega_0^{\lambda})
 (\omega_1^{\lambda}-\omega_2^{\lambda})},  \nonumber \\
&=& \int \frac{d^3 k}{(2\pi)^3 }\delta^2(p_1+p_2-k) 
 \sum_{\lambda}
  \left| \mathcal  M_{\lambda} \right|^2,
\end{eqnarray}
where we have used the notation $\mathcal 
 M_{\lambda} =(-ie) N_{k,\lambda} V^{\mu}  
\varepsilon_{\mu}^{\lambda}  $ 
 for the physical process
$\mathcal M_{\rm phys}(e^+e^-\to \gamma)$ shown in Fig. 2, and 
again we have introduced the normalization 
constant $N_{k,\lambda}=\frac{1}{\sqrt{2g\lambda 
 (\omega_1^{\lambda}-\omega_0^{\lambda})
 (\omega_1^{\lambda}-\omega_2^{\lambda})}} $.
The constant $N_{k,\lambda}$ can be understood in the following way:
 in the field expansion for usual photons, 
the normalization constant $\frac{1}
{\sqrt{2\omega_k}}$ 
comes from the delta
 $\delta^4(k^2)$ in four-momenta representation.
In our case the normalization constant is exactly the one 
 coming from $\delta^4({k^2+2g \lambda 
(n\cdot k)^3})$, and it can be verified that
 it has the correct limit when
 $g\to 0$, that is to say $N_{k,\lambda} \to \frac{1}{\sqrt{2\omega_k}}$.
Finally, we have
\begin{eqnarray}
2{\rm Im}[\mathcal M]= \int \frac{d^3 k}{(2\pi)^3 }
\delta^2(p_1+p_2-k) 
  \left| \mathcal  M_{\rm phys} \right|^2,
\end{eqnarray}
and therefore the unitarity condition is satisfied
 in this scattering process.
%%%%%%%%%%%%%%%%%%%%%%%%%%%%%%%%%%%%%%%%%%%%%%%%%%%%%%
\begin{figure} 
\centering
\includegraphics[
width=0.31\textwidth]{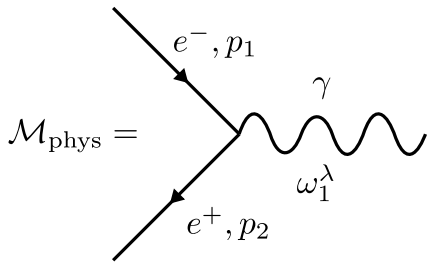}
\caption{Physical graph contributing to $\mathcal
 M_{\rm phys}(e^+e^-\to \gamma)$. }
\end{figure}
%%%%%%%%%%%%%%%%%%%%%%%%%%%%%%%%%%%%%%%%%%%%%%%%%%%%%%%%%%%%
%..............................................................................
\section{Conclusions}
%.............................................................................
In this work, we have studied whether perturbative
 unitarity in
a modified Lorentz-invariance violating QED theory with ghost states associated to
 higher-order time derivatives can be preserved. For this, 
we have focused on higher-order photons of the
Myers and Pospelov model
minimally coupled to standard fermions.

To summarize, we have identified two realizations 
of Lorentz symmetry breakdown in the Myers-Pospelov model
 where the dimension-five operators
lead to minimal modifications.
These occur when the breakdown is produced with a preferred 
four-vector in the timelike and spacelike  
directions.
For any other form of Lorentz invariance violation,
these dimension-five operators turn into 
higher-order time derivatives leading to  
ghost states that may produce the loss of unitarity of 
the $S$ matrix, thus undermining the probability interpretation of the theory.
With an explicit calculation we have verified that 
the unitarity condition 
in the process of electron-positron scattering at
tree-level order is satisfied. 
We have introduced a simplification by restricting 
only to physical degrees of freedom in the QED theory.
In this way we have bypassed the possible contribution 
of the usual ghost and longitudinal modes of standard electrodynamics.
The only ghosts we had to deal with were the ghosts coming from
the higher-order time derivatives of the theory.

The establishment of unitarity in our modified QED to order $e^2$ will require us to analyze 
more diagrams \cite{1loop}. Some of them contain loops 
where the ghosts can appear off-shell, thus, introducing an extra difficulty. 
Checking the unitarity condition 
to these orders will give us robust support in order to make physical predictions
in the theory.
%---------------------------------4+óp
\section*{Acknowledgments}
I want to thank Markos Maniatis and Luis Urrutia for reading the manuscript and for valuable 
comments on this work.
This work was 
partially supported by the Direcci\'on de Investigaci\'on de
la Universidad del B\'{\i}o-B\'{\i}o (DIUBB) Grant No. 123809 3/R.
%\vspace{150pt}
%...................................................................................................

\end{document}